**Discovery of Strange Kinetics in Bulk Material: Correlated Dipoles in $CaCu_3Ti_4O_{12}$**

A.M. Awasthi[*] and Jitender Kumar

Thermodynamics Laboratory, UGC-DAE Consortium for Scientific Research

University Campus, Khandwa Road, Indore- 452 001 (India)

## Abstract

Dielectric spectroscopy of $CaCu_3Ti_4O_{12}$ was performed spanning broad ranges of temperature (10-300K) and frequency (0.5Hz-2MHz). We attribute the permittivity step-fall to the evolution of Kirkwood-Fröehlich dipole-correlations; reducing the moment-density due to anti-parallel orienting dipoles, with decreasing temperature. Unambiguous sub-Arrhenic dispersion of the associated loss-peak reveals the prime role of strange kinetics; used to describe nonlinearity-governed meso-confined/fractal systems, witnessed here for the first time in a bulk material. Effective energy-scale is seen to follow thermal evolution of the moment density, and the maidenly estimated correlation-length achieves mesoscopic scale below 100K. Temperature dependence of correlations reveals emergence of a new, parallel-dipole-orientation branch below 85K. Novel features observed define a crossover temperature window connecting the single-dipoles regime and the correlated moments. Conciling known results, we suggest a fractal-like self-similar configuration of Ca/Cu-rich sub-phases; resultant heterogeneity endowing $CaCu_3Ti_4O_{12}$ its peculiar electrical behaviour.

---

[*] Corresponding Author: amawasthi@csr.res.in

The Perovskite-derivative $CaCu_3Ti_4O_{12}$ (CCTO) was reported in 2000 as a non-ferroelectric and non-relaxor material by Subramanian [1] et. al. Its crystal structure at room temperature was determined to be cubic, with lattice parameter 7.39Å and space group **Im3**. It exhibits colossal dielectric constant (CDC) of the order $\sim 10^4$ at room temperature, with weak *T*-dependence over the range 100-380K in both its single crystal and ceramic forms. Such properties make it very desirable for many technological applications like charge storage devices, memory elements, and microelectronics. Below nominally ~100K, anomalously huge and dispersive step-decrease in its dielectric constant is obtained, unlike the sharp resonance-maxima in relaxor ferroelectrics, with associated loss-peaks. From the neutron powder diffraction and high resolution X-ray diffraction though, no evidence of any structural transition has been found [1], although an antiferromagnetic (AFM) phase transition has been observed at 25K by magnetic susceptibility and specific heat measurements [2]. The very different electrical behaviours on either side of ~100K is an open question viz., whether it signifies a thermodynamic (order-disorder) or a kinetic (disorder-disorder/ergodicity-breaking) transition. So far, no corresponding features have been observed in thermodynamics and/or thermo-kinetics of CCTO near this temperature.

A controversy is long associated with CCTO regarding the mechanism responsible for its giant dielectric constant. Subramanian [1] et. al. originally suggested that CDC is due to the tilted-octahedral ($TiO_6$) crystal structure. Homes [3] et. al. performed optical and dielectric studies on single crystal of CCTO and attributed the peculiar electrical behavior to highly polarizable relaxational mode of dipolar fluctuations in nanosized domains. Zhu [4] et. al. related the huge dielectric constant of CCTO to the nanoscale disorder resulting from Ca-rich (insulating) and Cu-rich (conducting) phases, thereby enhancing the proximity effects. Large change of dielectric response by aliovalent cation dopants has been demonstrated [5-6], depending on the dopant and its particular-site occupancy within the unit cell. A detailed

STEM study on precisely site-imaging such dopants was performed by Choi [7] et. al. to understand and control the electrical behaviour. On the basis of impedance spectroscopy, Sinclair [8] and Adams [9-10] groups concluded that electrically heterogeneous CCTO ceramic contains semiconducting grains with insulating boundaries. This forms an internal barrier layer capacitance (IBLC), giving rise to the Maxwell-Wagner (M-W) mechanism, with two activation energies [8]; 0.08eV for the bulk and 0.6eV for the grain boundaries. The large *I-V* nonlinearity observed by Chung [11] et. al. was related to and confirms the IBLC picture.

The lack of long range electrical order [1], presence of nanoscale disorder [4], and huge nonlinearity [11] of CCTO indicate a major role of many-body interactions, leading to correlated/cooperative behaviour. Emergent systems such as CCTO are not adequately described in terms of conventional linear response, Debyean relaxation, and Arrhenian kinetics; they violate the superposition principle viz., their properties on whole are not simple sums of those of their elementary units. The framework of *strange kinetics* [12] (SK) is employed for such *complex systems* [13] (CS), with several distinguishing features; mesoscopic order being the primary one among them. Prominent character of CS with SK is the fractional power-law-exponent forms of their spatio-temporal correlation functions, borne of scale-invariance/self-similarity. The IBLC/M-W mechanism for the CDC above ~100K and the decreasing moment-density for its sharp fall below are phenomenological descriptions. Identification of exact material sub-phases, their spatial-configuration, and building-blocks responsible for the peculiar electrical response in CCTO remain unexplored. The present study addresses these important issues from the perspective of strange-kinetics [12] as observed, based on a broadband analysis of our results in terms of mid-range dynamic correlations and material-substructure. We evaluate kinetic & spectral parameters that affirm the absence of thermodynamic/kinetic phase transition. Our findings present evidence of dipole-interactions and fractal-type self-similar microstructure realized from phase-segregated constituents.

The ceramic CCTO samples were prepared from high purity (99.99%) powders of $CaCO_3$, CuO, and $TiO_2$ by the conventional solid state route. The pelletized samples (10mm diameter and 1-3mm thick) were sintered at 1100°C and were silver-coated for good electrical contacts for dielectric measurements. XRD of the samples has been done by RIGAKU rotating anode powder diffractometer with Cu-$K_a$ radiation ($l$ = 1.54Å). Dielectric measurements over 10K-room temperature were performed using NOVO-CONTROL (Alpha-A) High Performance Frequency Analyzer with 1V ac signal covering 0.5Hz to 2MHz. Specific heat data were obtained using semi-adiabatic method over 4.2-200K.

Our dielectric spectroscopy data on CCTO exhibit high dielectric constant $e'\sim 10^4$ near room temperature (low frequencies) and its $T(w)$-shifted but $w$-independent plateau over 100-200K, with frequency-dispersive sharp drop (fig.1a) typically below ~100K scale. We also observe small but clear non-dispersive step-down changes in $e'(T)$ at the antiferromagnetic ordering temperature $T_N$ = 25K, reported previously by Grubbs [6] et. al. in their Fe/Nb-doped CCTO only, and related by them to magneto-electric coupling. Imaginary permittivity $e''$ shows dispersive loss peaks (fig.1b) over 40-250K interval, with systematically varying width and height. These features are in common with the reported results, both on polycrystals as well as single crystal [3]. We also show our large-variation $C_p$ data (fig.1a inset) marking the $T_N$, without any corresponding feature near the elusive ~100K.

As the long-range electrical/structural ordering has been clearly ruled out in CCTO [1], its peculiar thermo-spectral dielectric behaviour mandates exploring the signatures of mesoscale dynamical organization. This is borne of Ca/Cu site-occupancy disorder; however, being dynamic, it does not lead to frozen (real space) polar nano regions (PNR's), as in the disorder-quenched relaxor ferroelectrics. The mid-range dynamical correlations have been examined by the Kirkwood-Fröehlich factor [14-15]

$$B(T) = T[e_{st}(T) - e_{\infty}]\frac{2e_{st}(T) + e_{\infty}}{3e_{st}(T)} = \frac{\langle M^2 \rangle}{3e_0 k_B V}, \tag{1}$$

expressed in terms of the static and high-frequency susceptibility, and the mean-square moment in a correlation volume $V$ containing $N$ dipoles that relax collectively ($\langle M^2 \rangle \neq N\mathbf{m}^2$). The positive (negative) slope $dB/dT$ signifies rather/time-averaged anti-parallel (parallel) orientation-correlation of the neighboring dipoles. The antiparallel arrangement of local dipole moments (due to $TiO_6$) has also been mentioned by Onodera [16] et. al., on the basis of corner-sharing of the $CuO_4$ square planes with the surrounding tilted octahedra. Temperature dependence of the K-F factor for CCTO shown in fig.2 generically resembles that reported for water in porous silica glass [15]. Monotonic increase of $B(T)$ for CCTO indicates the rise of mean-square moment density (MSMD). De-correlation of relatively anti-parallel configuration of neighboring dipoles ($\langle M^2 \rangle \leq N\mathbf{m}^2$) to attain its free dipole value (=$N\mathbf{m}^2$) has also been alluded to by Homes [3] et. al. The broad asymptotic saturation of $B(T)$ signals correlations' slow decay ($N \to 1$) at high temperatures, and seems to be consistent with the step-rise of permittivity $e'(T)$ followed by its plateau.

Noticing that the maxima of the loss-peak in $e''$ track the mid-points of the $e'$-step-rise, the observed dispersion itself defines the frequency-scale $w_{corr}(T)$ of the dipolar-correlations; its increase with temperature implying a decreasing correlation size $V_{corr}(T)$. For the soft matter in confined geometries, this sizescale is set by the enclosure's porosity. As regards the electrical characteristics of CCTO as a complex system [13], the fractal disorder in Ca/Cu site-occupancy constrains the spatial extent over which the dipolar dynamics is correlated, inhibiting the establishment of a long range order. An idea of the change in the MSMD can be obtained by estimating its extremes versus temperature. To this effect, we made a sigmoid-function fit on our $B(T)$ (evaluated for the limited $T$-interval), and extrapolated the same over a wide temperature range (fig.2, upper inset, left-axis). Ratio of the high- and low-temperature K-F factors equals that of the fully-free and maximally-correlated MSMD's:

$$\frac{B(T_{hi})}{B(T_{lo})} = \frac{\langle M^2/V \rangle_{free}}{\langle M^2/V \rangle_{corr}} = 13.5, \qquad (2)$$

remarkably matching with the magnitude-change witnessed in the permittivity step-rise.

To further examine the correlations, we plot the slope $dB/dT$ (fig.2 upper inset, right-axis) of the sigmoid-fit to the K-F factor, showing a maximum at ~85K. What could possibly decrease an otherwise increasing slope (intuitively expected and still positive), below certain temperature? To comprehend this somewhat anomalous behaviour, we need to look into all possible nearest-neighbor dipoles' pairing-configurations. In the CCTO structural-cell we notice that out of the six $TiO_6$ blocks nearest to a randomly-selected one, four are laterally and two are axially positioned relative to it. Now, the stronger interaction between sidewise-placed dipoles would manifest at a higher temperature, compared to the weaker one between serially-placed dipoles. Remember that the positive (negative) slope-values of $B(T)$ signify anti-parallel (parallel) nearest-neighbor dipole-correlations. Thus, it transpires that while the four (transverse) neighbors prefer relatively anti-parallel orientation, apparently the two (axial) neighbors with weaker interaction opt for the parallel one; causing the observed $B(T)$ to decrease progressively slowly below the "correlation-branch" temperature $T_{cb}$, where the 'axial' parallel-correlations may debut.

For quantitative estimates of the degree of correlations, The K-F factor is written [14-15]

$$B = B_0 \left[ 1 + \sum_n z_n \langle \cos\mathbf{q} \rangle \right], \qquad (3)$$

where $\mathbf{q}$ is the angle between a dipole moment and its ($z_n$) nearest neighbors (having the same interaction); summation extending over different types of the latter. For our case here, above 85K, $z_1 = 4$ (transverse neighbors, with their $\langle\cos\boldsymbol{q}_{\uparrow\downarrow}\rangle \leq 0$) solely contribute to the summation in *B*, while at lower temperatures $z_2 = 2$ (axial neighbors, with their $\langle\cos\boldsymbol{q}_{\uparrow\uparrow}\rangle \geq 0$) would also contribute. To evaluate the correlation effects versus temperature, we take the upper saturation value of the K-F factor-fit as (*T*-invariant) $B_0$ (no correlations, $\langle\cos\boldsymbol{q}\rangle \equiv 0$), and take its extrapolation below 85K with continued power-law behavior, to keep representing the anti-parallel correlations. The resultant functions are plotted vs. temperature in fig.2 (lower inset) as $(1+4\langle\cos\boldsymbol{q}_{\uparrow\downarrow}\rangle)$ (left-axis), causing a maximum reduction of the MSMD by a factor of over 400, and as $(2\langle\cos\boldsymbol{q}_{\uparrow\uparrow}\rangle)$ (right-axis), enhancing MSMD by a maximum of ~7%. In particular, at room temperature (300K), the anti-parallel correlations reduce MSMD to ~62% of their high-*T*/free-dipoles value, and to ~12% of the same at 44K (lowest temperature of our dielectric dispersion peak). Contrastingly, at 44K the parallel correlations (emerging below 85K) cause an increase of the MSMD by a mere 0.67%.

Another manifestation of the dynamic dipole-correlations is seen in the anomalous dispersion kinetics obtained from the loss-peak maxima. Arrott plot of the effective times $\boldsymbol{t}_{\text{eff}} \approx 1/2\pi f_p$ versus $T^{-1}$ is sublinear over a broadband (60dB) of $\boldsymbol{t}_{\text{eff}}$ values (Fig.3, main panel); similar forms reported e.g., as the polaronic relaxation in $SrTiO_3$ [17], debated in the low-*T* dielectric dispersion of $CdCr_2S_4$ [18-20], and referred to as the interaction effects in CCTO [21]. While ruling out pure relaxation of thermally robust units (Arrhenian kinetics [22-23]), this non-linearity does not relate to the segmental dynamics of cooperatively rearranging moments [24-25], which leads to super-Arrhenicity (Vogel-Fulcher behaviour/superlinear Arrott plot) for glasses (force-fitting in our case gives a negative/unrealistic $T_{\text{VF}}$). Moreover, as no structural or electrical reordering occurs over this range, a first-order phase transition is inadmissible. Spectral character of the dispersion peak discussed later confirms that this anomalous (sub-Arrhenius) behaviour cannot be an artifact of the conductivity effects, as argued for the case of $CdCr_2S_4$ in [20], to rule out the forced Vogel-Fulcher fit suggested in [19] with unphysical $T_{VF}$. Besides this wide-range sub-Arrhenicity of its dielectric-loss kinetics, CCTO possesses not only one of the largest known $d(\log I)/d(\log V)$; $\boldsymbol{a}_{CCTO} \geq 900$ [11] >> $\boldsymbol{a}_{ZnO} \cong 30\text{-}80$ [26], but also displays appreciable local variation [27] in its *I-V* nonlinearity, within the grains. The latter confirms that the electrical inhomogeneity, also characterizing the single crystals [4], is intrinsic. These evidently suggest that the insulating and conducting regions in CCTO are intertwined into fractal configuration. The self-similar character of fractal structures also explains the broad plateau in stepped-up $\boldsymbol{e}'_w(T)$; as convoluted response from distributed spatio-temporal-energetic-scale entities, confirmed by para-Debyean spectral widths of dielectric loss-peaks (see later, fig.4). Nonlinearity and self-similarity definitely endow CCTO qualification as a complex system [13], to be appropriately described by the strange kinetics [12].

Our dispersion data excellently fits over the full (6-decade) *t*-range by a combined exp-linear function $\ln \boldsymbol{t} \sim [\alpha T^{-1} - \beta \exp(-\gamma T^{-1})]$. The strange kinetics [13,28] observed here is directly related to dipole-correlations, also reported e.g., by Chen [21] et. al. as quadratic (power-law ≡ short-range) correction to the $\sim T^{-1}$ argument of the otherwise Arrhenius exponent. Our exp correction (polynomial ≡ medium-range) corresponds to a hierarchy of dipole-interaction terms. Similar form (albeit, with opposite sign) has been reported in connection with the confinement kinetics in KTN:Cu crystal [29-30], identifying two (reorientation and defect-formation) activation energies. Clearly, global (constant) activation energy cannot be extracted here, as is possible in the Arrhenian and VFT cases. Nonetheless, one can define an "effective" barrier energy function, as the slope $E_{eff}(T) = d\text{Ln}\boldsymbol{t}_{eff}/d(1/T)$. Thus-obtained derivative (of the exp-lin fit) is shown in fig.3 (inset); understandably, values in different *T*-regimes correspond to those reported in the literature, obtained by linearizing small sections of the full nonlinear curve [3,6,31-32]. It is of interest to examine the simple limiting behaviours of this peculiar kinetics. Using scaling of the correction term ($\gamma \cong 76K$) our fit extrapolates to bare and renormalized asymptotic Arrhenian forms, as shown in fig.3. Consequently, effective energy saturates to vastly different values: $E_{eff}(0) \sim 8.8$meV as $\exp(-gT^{-1}) \xrightarrow{T \to 0} 0$, and $E_{eff}(T \gg 76K) \sim 170$meV as $\exp(-gT^{-1}) \xrightarrow{T \gg 76K} 1 - gT^{-1}$. Different $e'(T)$ at these $T$-extremes thus clearly refer to distinct moment-densities.

The evaluated $E_{eff}(T)$ does show tendencies towards asymptotic saturation; constant $E_{eff}$ at higher (lower) temperature are more akin as due to the bare (correlated) dipoles. A rather smooth inflexion point is recognizable at exactly the crossing (76K) of the two Arrhenian asymptotes. Dispersive loss-peak and step-rise in $e'(T)$ (fig.1a), as well as saturation of the effective energy (fig.3 inset) evidence both a rising frequency-scale $f_p(T)$ and the allied reducing length-scale ($l_{corr}$, most reasonably $\propto f_p^{-1/3}$) of correlations, with increasing $T$. Decrease of $l_{corr}$ to the nearest-neighbor dipole-dipole distance (effective-size of free-dipoles) would then signal a demise of correlations. Our exp-lin fit gives the fastest response time ($t_0$ ~0.47ns $\equiv$340MHz) as the limit for frequency-dependence to be observable, and for the $e'_w(T)$-step/loss-peak to disappear. Literature report by Lunkenheimer [33] group provides evidence of the kind at frequencies $\geq$ 200MHz, just above the room temperature.

We digress now to examine the allied spectral anomalies of the intrinsic response, via peak-scaled losses { $e''(w/w_m, T)/e''_m(T)$ } at various temperatures, plotted in fig.4. Dispersion of the curves at their lower-frequency ends is due to the conductivity effects (ignorable to within -4dB (at/below 60K) and to within -6dB (at/above 200K) of the peak-maximum); unaffecting the peaks' intrinsic spectral character, which clarifies its above-analyzed sub-Arrhenic kinetics of artifacts. Noteworthy features distinguish our ensemble from overlapping master curves that represent generic/typically-Debyean relaxation, governed by single energy- & time-scale [34-35]. Non-universality of the scaled loss-peaks observed here corresponds to (*T*-variant) energy scale and distribution of relaxation times [36-37]; a natural character of intrinsic fractality. This 'broken symmetry' para-Debyean spectral feature against temperature is but the *w*-domain manifestation of non-Arrhenic kinetics witnessed in the *T*-domain, traced to thermal evolution of spatio-spectral correlations and their effective energy scale. The correlated nature of dipoles has also been confirmed by the Dissado-Hill [38-39] power-law spectral fits of our permittivity, which also reconciles with the intrinsic self-similarity feature [40]. Moreover, the nanoscale-disorder/fractality of CCTO manifests signature in its thermal conductivity conforming to those of amorphous/glassy materials; finite scope of the present article necessitates reporting the full implications of these analysis/results elsewhere [41]. The loss-peak maxima $w_{mx}(T)$ represent the frequency scale of these correlations, since by causality [36-37,42-43]

$$e'' \approx - (p/2)\, \partial e'/\partial \ln(w), \qquad (4)$$

and $e'(T_{mx})$ is middle (inflexion-point) of the step-rise, latter signaling the correlation effects. A discontinuous jump in the loss-peak-asymmetry {(LW-UW)/(LW+UW)} at 91K (fig.4 inset) may be related to the onset of parallel-orientation dipole-correlations, causing a maximum in ($dB/dT$) at 85K. Fractional power-law exponent of the loss-peak spectral-width (FWHM $\propto T^{-3/4}$) below 100K, rolling-off towards a low level-value (Fig.5, left *y*-axis) confirms the approach to bare Arrhenicity, consistent with the high-*T* asymptotic kinetics. Product of the loss-peak height and FWHM provides the total (*w*-integrated) dielectric losses

$$W = \int e'' d(\ln w) = e''_{max} \log(w_{uh}/w_{lh}), \qquad (5)$$

shown versus temperature in fig.5, right *y*-axis. Clear maximum evident in *W* at exactly 100K may signify highest electrical heterogeneity (lengthscale-distribution/spatial-fractality); first evidence of a specific feature seen at this nominally-quoted temperature.

Figure 5 (lower inset) shows the power-law covariance of FWHM & frequency-scale $\left(w_{corr} \propto \{1+4\langle\cos q_{\uparrow\downarrow}\rangle\}^{47/5}\right)$ with the degree of (anti-parallel) dipolar-correlations $(\langle m_i m_{i+1}\rangle)$. This in turn implies that the electrical susceptibility $c = e$ -1 (related to the latter in Kubo formalism [44], based on the fluctuation-dissipation theorem) and $e^* \propto w^p$, as only too evident from the log-log plot in fig.4, and confirmed via the Dissado-Hill spectral-fits of our measured permittivity [41]. This allows an estimation of the correlation length-scale (fig.5, upper inset), with $l_{corr} \sim T^{-3.24}$ dependence below 100K (solid line fit); assuming rather reasonably $l_{corr} = V_{corr}^{1/3} \sim w_{corr}^{-1/3}$, and equating $V_{corr}$ at high-$T$ extremes to the effective volume per $TiO_6$ octahedron (2octs. per unit cell; $l_{corr} \xrightarrow{\langle\cos q_{\uparrow\downarrow}\rangle\to 0} \sim a/4 \cong 1.83 \text{Å}$). It is obvious that (*a*) at room temperature the correlations extend just beyond the unit cell--- $l_{corr}^{RT} \sim 8\text{Å}$, (*b*) the major evolution of $l_{corr}$ takes place only below ~100K, and (*c*) the mesoscopic length of $l_{corr}$ ~155 nm is realized at 44K. An emergent temperature window hosting several spectro-kinetic features thus represents a crossover from bare- to correlated-moments regimes. To seek commonality between our bulk CCTO and the complex systems, both sharing the strange kinetics, we envision the Ca-rich (insulating) and Cu-rich (conducting) phases [4] as fractally intertwined. This perspective admits the establishment of mesoscopic correlation lengths and fractally self-similar IBLC-substructures for high-dielectric response, as causing the peculiar *e′*-step-rise and CDC in CCTO.

To summarize, our broadband dielectric study of $CaCu_3Ti_4O_{12}$ mandates strange kinetics as the first instance of its detection in the bulk. Signatures of intrinsic nonlinearity as sub-Arrhenic kinetics, and of fractality in huge permittivity & its broad plateau accord CCTO an anomalous complex-material attribute. The effective energy and time scales here correspond to the Kirkwood-Fröehlich correlations with 'anti-parallel' nearest-neighbor dipole-orientation. We witness the emergence of a parallel-dipole-orientation correlation branch below $T_{cb}$ = 85K. Exactly at the long sought-after 100K, maximum in the integrated losses signifies electrically most heterogeneous state, with widest distribution of lengthscales. Our effective lengthscale estimates of the dynamic correlations indicate a mesoscopic substructure. The crossover evolution of free dipoles at high temperatures into dynamically correlated moments at low-temperatures excludes a thermodynamic or kinetic phase transition. Compatible with the current knowledge, our study points to a fractal sub-configuration of Ca- and Cu-rich regions. The insight can be utilized to tune the CDC and characteristic-temperatures of CCTO; by enhancing the Ca/Cu site-occupancy disorder via chemical, thermal, mechanical, or electrical means, and to design advanced generation of CDC materials. Precision spectro-microscopy ought to divulge the full spatio-dynamic features, and map the electrical topography in $CaCu_3Ti_4O_{12}$.

## Acknowledgements

The authors thank R. Rawat for providing the specific heat data. P. Chaddah and A. Gupta are thankfully acknowledged for their support and encouragement.

**Figure Captions**

**Figure 1.** Real and imaginary permittivity of $CaCu_3Ti_4O_{12}$ versus temperature across 0.5Hz-2MHz. Anomalous *T*-dependence and dispersion of *e'* (*e''*) is due to correlation & relaxation effects. The AFM transition at 25K is detected in permittivity. Relative invariance (vs. *w*, *T*) of *e'*(*T*)-plateau signify intrinsic fractality; convoluted outcome of distributed spatio-temporal-energetic scales of responding entities (see text). For higher frequencies (≥340MHz, see text), the $e'_w(T)$-step is expected to vanish; so observed [33] above 200MHz. Inset-- besides the $T_N$-peak, $C_p(T)$ displays no corresponding features peculiar to *e**(*w*, *T*).

**Figure 2.** The Kirkwood-Fröehlich correlation factor *B*(*T*) monotonically increases, signifying relatively anti-parallel orientation of nearest dipoles [14-15], and tends to saturate at high temperatures (to the free-dipoles value). Upper inset-- (left-axis) Boltzmann sigmoidal fitted onto *B*(*T*) estimates a maximum 13.5 times increase over the 10K roll-off, of the mean-square moment density; (right-axis) derivative (*dB/dT* > 0) of the sigmoid-fit with a maximum signals the emergence of (weaker) parallel-orientation correlations at the 'correlation-branch' temperature $T_{cb}$ = 85K. Lower inset-- *T*-dependences of the two dipole-dipole correlations: (left-axis) anti-parallel orientations $(1+4\langle \cos q_{\uparrow\downarrow}\rangle)$; (right-axis) parallel orientations $(2\langle \cos q_{\uparrow\uparrow}\rangle)$ (see text).

**Figure 3.** Arrott plot showing unambiguous sub-Arrhenic (strange) kinetics associated with the peculiar dielectric dispersion in CCTO. The exp-lin fit $\{\ln t \sim [\alpha T^{-1}-\beta \exp(-\gamma T^{-1})]\}$ reflects dipolar correlations, with both relaxation and evolution of dynamic clusters contributing to the kinetics. Inset-- effective energy evaluated from the slope $E_{eff} = d\ln t/dT^{-1}$ has a wide-sigmoidal shape, its high (low) temperature bare (renormalized) Arrhenius saturation represented by linear asymptotes in the main panel (see text).

**Figure 4.** Normalized imaginary permittivity versus peak-scaled frequency. Absence of a master curve is consistent with the non-Arrhenic kinetics over the observed *w*-*T* domain of dispersion. Para-Debyean peak-widths signify interacting moments, having temporal-energetic distributions [36-37]. Inset-- asymmetry of the loss-peak with discontinuous down-jump at 90K and a shallow minimum at ~120K mark major changes in the evolution/relaxation of dynamically-correlated dipoles (see text).

**Figure 5.** FWHM of the loss peak (left-axis) with fractional power-law exponent ($\sim T^{-3/4}$) below 100K and tendency to roll-off at high temperatures signifies reducing dipole-correlations. Spectrally-integrated dielectric losses (right-axis) have a maximum at exactly 100K, related to the mesoscopic heterogeneity. Lower inset--- co-variations of the correlation frequency $\left(w_{corr} \propto \left\{1 + 4\langle\cos q_{\uparrow\downarrow}\rangle\right\}^{47/5}\right)$ (left-axis) and FWHM (right-axis) with the anti-parallel correlation function may relate to the power-law *w*-dependence of the permittivity (see text). Upper inset--- effective correlation length ($l_{corr} \sim T^{3.24}$ below 100K), assuming $l_{corr} \xrightarrow{\langle\cos q_{\uparrow\downarrow}\rangle \to 0} \sim a/4$ (~ linear-size of a $TiO_6$ octahedron) (see text). At room temperature, the correlation-effects are barely expected $\left(l_{corr}^{RT} \sim 8\text{Å}\right)$; however, $l_{corr}(44K) = 155$nm justifies the peculiar dielectric behavior of CCTO across ~100K as a mesoscopic phenomenon.

**Fig.1**

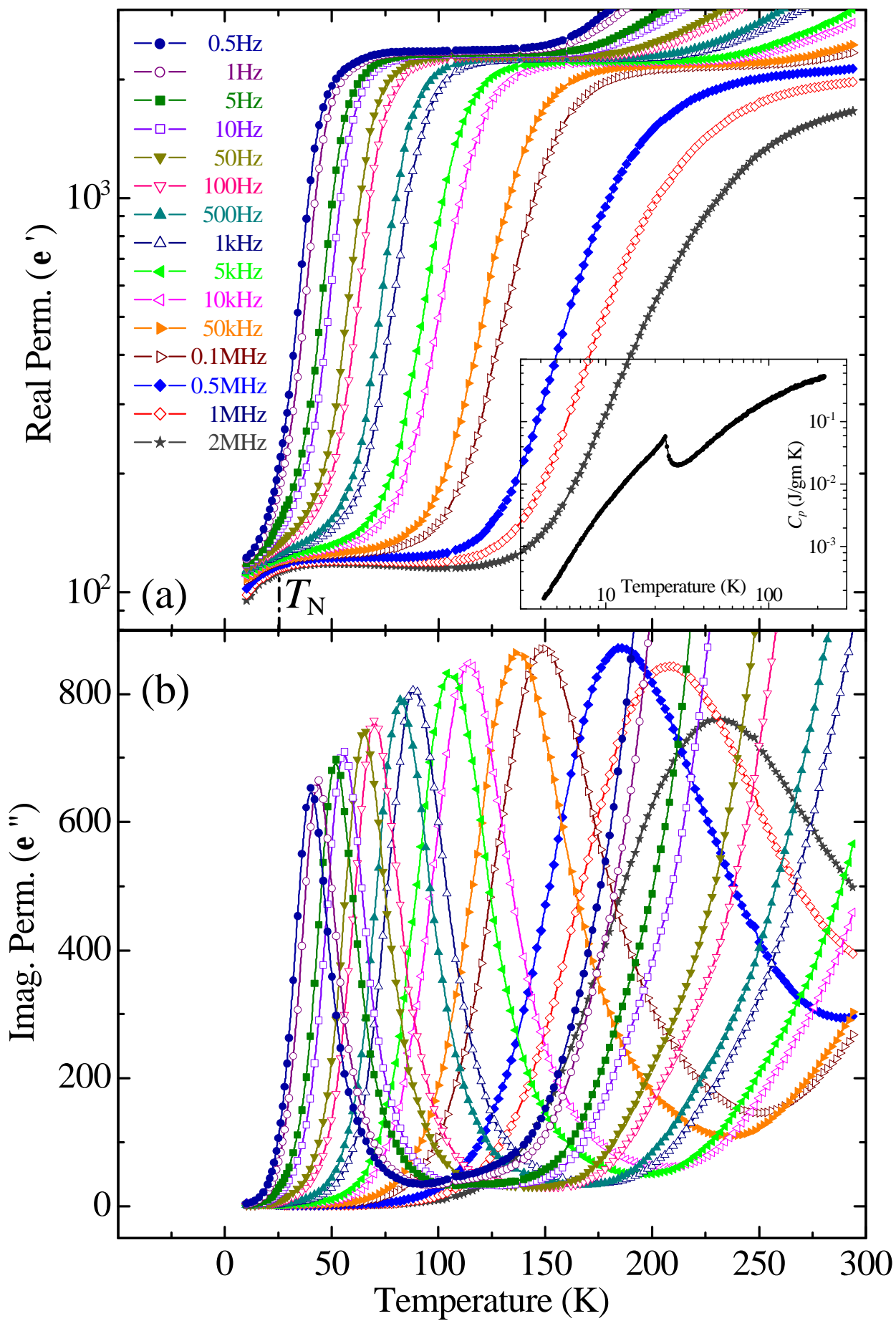

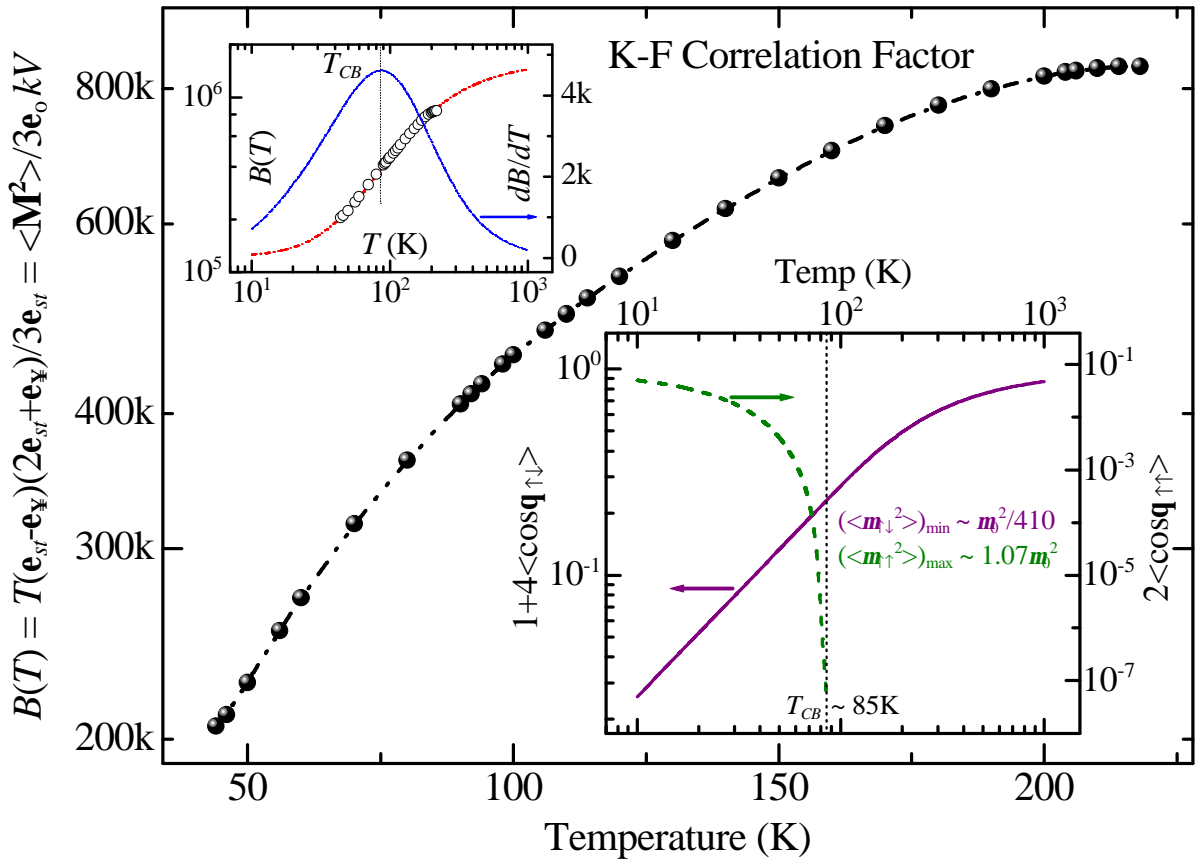

Fig.2

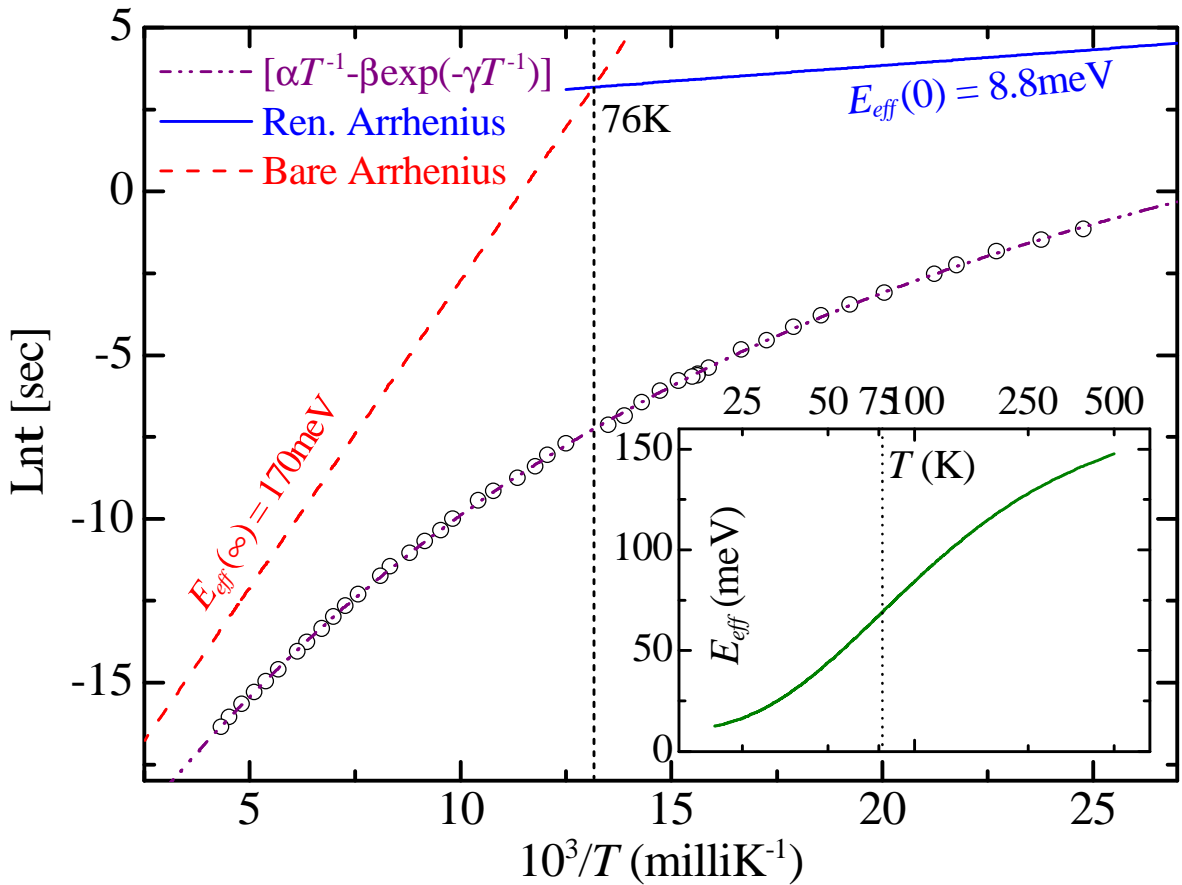

Fig.3

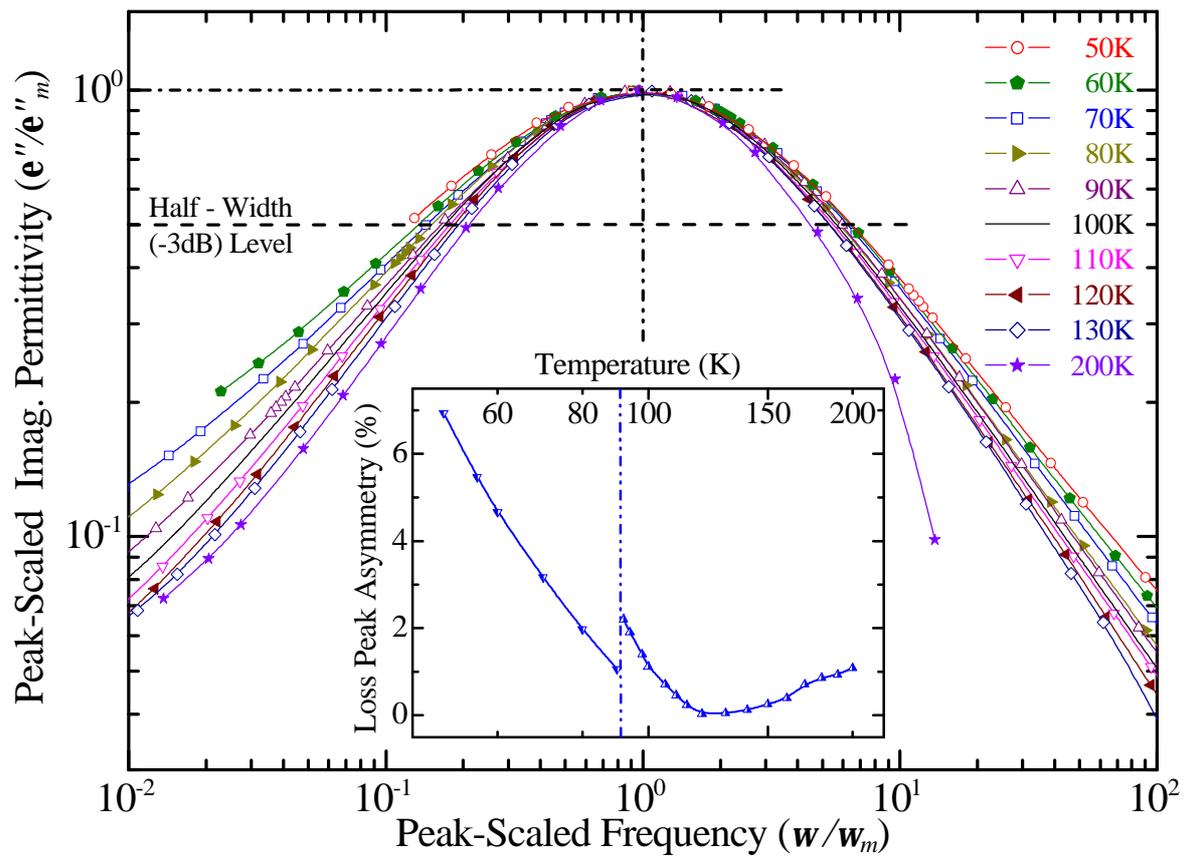

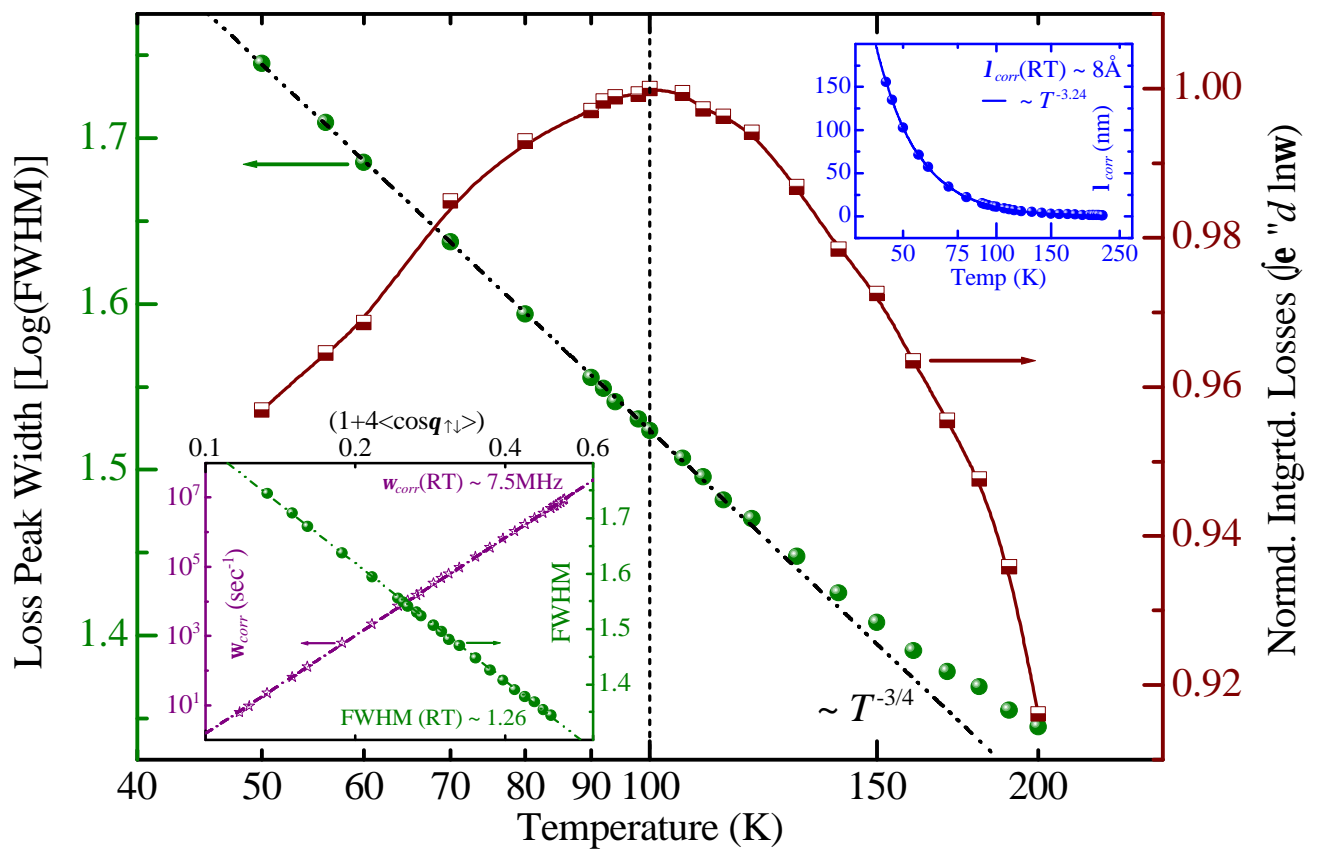